\documentclass[aps,pra,preprint,showpacs]{revtex4}
\usepackage{hyperref}
\usepackage{setstack}
\usepackage{amscd}
\usepackage{amssymb}

\usepackage[mathscr]{eucal}
\usepackage{dsfont}
\newcommand{\1}{\mathds{ 1}}
\newcommand{\ket}[1]{| #1 \rangle}
\newcommand{\bra}[1]{\langle #1 |}

\usepackage[utf8x]{inputenc}

\newcommand{\zukowski}{{\ifmmode \dot{Z}\else \.{Z}\fi{}ukowski} }

\begin{document}
\title{Bound entanglement helps to reduce communication complexity}
\date{\today}

\author{Michael \surname{Epping}}
\email{epping@thphy.uni-duesseldorf.de}
\affiliation{Institut f\"{u}r Theoretische Physik III, Heinrich-Heine-Universit\"{a}t D\"{u}sseldorf, Universit\"{a}tsstr. 1, D-40225
D\"{u}sseldorf, Germany}
\author{\v{C}aslav \surname{Brukner}}
\affiliation{Vienna Center for Quantum Science and Technology (VCQ) and Faculty of Physics, University of Vienna, Boltzmanngasse 5, A-1090
Vienna, Austria}
\affiliation{Institute of Quantum Optics and Quantum Information (IQOQI), Austrian Academy of Sciences, Boltzmanngasse 3, 1090 Vienna,
Austria}


\begin{abstract}
We present a simple communication complexity problem where three parties benefit from sharing bound entanglement. This demonstrates
that entanglement distillability of the shared state is not necessary in order to surpass classical communication complexity.
\end{abstract}

\maketitle

\section{Introduction}
Quantum Information studies communication or computation schemes which allow more efficient solutions when
considering the laws of quantum theory instead of those of classical physics. In this research field, entanglement has proven to be
 a beneficial resource and
many applications make use of maximally entangled states~\cite{nielsen2000quantum}. As these states are important for such applications,
methods have been developed to create one maximally entangled state out of several copies of less entangled states using local
operations and classical communication (LOCC)~\cite{PhysRevLett.76.722}. This process is called entanglement distillation. Entangled states
that allow for the creation of a maximally entangled state by LOCC in at least one bipartition of the composite system are called
distillable. States which are entangled but not distillable are called bound entangled~\cite{PhysRevLett.80.5239}.\\ 
Bell inequalities are constraints on probabilities for local measurements, which are satisfied by local hidden variable
theories~\cite{Bell64,Peres99}. However, they are not satisfied by quantum mechanics. Entangled states that violate a Bell inequality are
called nonlocal. There exist (mixed) entangled local states, i.e. states which do not violate any Bell
inequality~\cite{PhysRevA.40.4277}. Yet, it was shown that all entangled states, including bound entangled ones, violate a Bell inequality
when combined with another state which on its own cannot violate the same Bell
inequality~\cite{PhysRevA.86.052115}.\\
Every distillable state may be transformed into a nonlocal state using only LOCC, but not every nonlocal state is
distillable. This was found recently by giving an explicit example of a nonlocal bound entangled state~\cite{PhysRevLett.108.030403} 
(strengthening previous results~\cite{PhysRevLett.87.230402,PhysRevA.74.010305,Acin2001} to fully bound entangled states, see below for the
definition of fully bound entangled states). Even though no pure entanglement can be distilled from bound entangled states they constitute
an useful resource in quantum information protocols. These are entanglement activation~\cite{PhysRevLett.82.1056,MasanesExtracted},
enhancement of the teleportation power of some other state~\cite{PhysRevLett.96.150501}, quantum steering~\cite{BrunnerSteering}, quantum
data hiding~\cite{Chattopadhyay2007273} and quantum key distribution~\cite{4529274}. The last two tasks are ``classical'' in the sense that
they can be stated outside the framework of quantum theory. Quantum theory can then enable advantages in comparison to how the tasks can be
performed on the basis of classical laws. In this paper we consider another task of this type - communication complexity - for which we show
that bound entangled states can provide advantage over all possible classical solutions. This task allows to quantify the advantage of the
bound entangled states with respect to classical resources of shared (classically) correlated bit strings. Communication Complexity studies
the amount of information that must be communicated between distant parties in order to calculate a function of arguments which are
distributed among the parties~\cite{Kushilevitz:1996:CC:264772}. We consider a similar question: If the parties are restricted to
communicate only a given amount of information, what is the highest possible probability for them to estimate the value of the function
correctly?\\
It is well known, that nonlocal states can be useful in such a task~\cite{PhysRevLett.92.127901}. Here we give a surprisingly simple
example illustrating the fact, that this includes even fully bound entangled states. 
\section{A general quantum communication complexity scheme}
We will make use of a generalization of the quantum communication complexity scheme introduced in Ref.~\cite{PhysRevLett.92.127901} to more
than two bits input per party. Consider the situtation where $n$ parties labelled $1$ to $n$ are spatially separated. Let us assume an 
inequality of the form
\begin{equation}
 \sum_{x_1,...,x_n=0}^{2^m-1} g(x_1,...,x_n) E(x_1,...,x_n) \leq B, \label{eq:bellineq}
\end{equation}
where the coefficients $g(x_1,...,x_n)$ and the local hidden variable bound $B$ are real numbers and $E(x_1,...,x_n)$ is the correlation
function of a measurement for the choice of measurement setting $x_i$ by each party $i$. The correlation function can be expressed as 
$E(x_1,...,x_n)=P(a_1...a_n=1|x_1,...,x_n)-P(a_1...a_n=-1|x_1...x_n)$, where $a_i=\pm 1$ is the measurement result of observer $i$. 
We call inequality~\ref{eq:bellineq} a Bell inequality, if it can be violated by a value $S>B$ using quantum mechanical expectation values.
Following the idea of Ref.~\cite{PhysRevLett.92.127901} we introduce a quantum communication complexity problem associated with this Bell
inequality. Each party $i$ receives one bit $y_i\in\{-1,1\}$ and $m$ bits $x_i\in\{0,1,...,2^m-1\}$ unknown to all the other parties. The
two possible values of $y_i$ occur with equal probability while the values of $x_i$ follow the probability distribution
\begin{equation}
 Q(x_1,...,x_n)=\frac{|g(x_1,...,x_n)|}{\sum_{x_1',...,x_n'=0}^{2^m-1} |g(x_1',...,x_n')|},\label{eq:Q}
\end{equation}
which is fixed beforehand and known to all parties. Their common task is to output the value of the function
\begin{equation}
f(y_1,...,y_n,x_1,...,x_n)=\prod_{i=1}^n y_i \mbox{sign}\left[g(x_1,...,x_n)\right].
\end{equation}
The parties will not evaluate the function correctly with certainty. The aim is to maximize the probability of successful evaluation. Each
party is allowed to broadcast a single bit of information to its fellow parties. It is required that all parties broadcast the bit
simultaneously (in this way the communicated bit of one party does not depend on the broadcasted bits of others, but only on the local
input). Afterwards one of the parties is asked to output the value of the function. We consider two different protocols. In the classical
protocol the bit $s_i$ sent by party $i$ could be in general, any function of $y_i$ and $x_i$. However it was shown
in Ref.~\cite{masterarbeit}  (analog to Ref.~\cite{PhysRevLett.92.127901}) that in the optimal classical protocol $s_i=y_i a_i(x_i)$ where
$a_i(x_i)$ is an appropriate chosen function $\{0,1,...,2^m-1\}\rightarrow\{-1,1\}$ and the best guess is given by
\begin{equation}
 A(y_1,...,y_n,x_1,...,x_n)= \prod_{i=1}^n y_i a_i(x_i).\label{eq:answerfunction}
\end{equation}
Intuitively one can understand this in the following way. Opposite values of any $y_i$ lead to opposite values of the function $f$. Missing
a single $y_i$ would completely destroy the information about the result. Therefore it is crucial to communicate $y_i$ in a way that allows
to reconstruct the product of all the $y_i$'s. In the quantum protocol $a_i(x_i)$ is replaced by the measurement result $a_i$. Each party
$i$ chooses one out of $2^m$ possible measurement settings according to the input $x_i$ and sends $y_i$ multiplied by the measurement
result $a_i$. The best guess is then again given by Eq.~\ref{eq:answerfunction}.\\
The probability of success of the protocol, i.e. the probability for $A(y_1,...,y_n,x_1,...,x_n)$ to equal $f(y_1,...,y_n,x_1,...,x_n)$ can
be written as
\begin{equation}
P(A=f)=\frac{1}{2}\left[1+(f,A)\right]
\end{equation}
using the weighted scalar product
\begin{equation}
(f,A)= \sum_{y_1,...,y_n=\pm 1} \sum_{x_1,...,x_n=0}^{2^m-1}  \frac{1}{2^n} Q(x_1,...,x_n) f(y_1,...,x_n)
A(y_1,...,x_n).
\end{equation}
Inserting $Q$, $f$ and $A$ gives the probability of guessing correctly 
\begin{equation}
 P_C=\frac{1}{2}\left(1+\frac{B}{\sum_{x_1,...,x_n=0}^{2^m-1} |g(x_1,...,x_n)|}\right) \label{eq:PC}
\end{equation}
in the classical protocol and
\begin{equation}
 P_Q=\frac{1}{2}\left(1+\frac{S}{\sum_{x_1,...,x_n=0}^{2^m-1} |g(x_1,...,x_n)|}\right) \label{eq:PQ}
\end{equation}
in the quantum case. 
\section{Bound entanglement as a resource}
We now come to the explicit example. We choose $n=3$, so there are three separated parties. They share the state 
\begin{equation}
\rho=\sum_{i=1}^4 p_i \ket{\psi_i}\bra{\psi_i}\\
\end{equation}
with $p_1=0.0636039$, $p_2=p_3=0.273734$, $p_4=0.388929$ and
\begin{eqnarray*}
 \ket{\psi_1}&=&0.183013\ket{000}-0.408248\left(\ket{001}+\ket{010}+\ket{100}\right)+0.683013\ket{111},\\
 \ket{\psi_2}&=&-0.344106(\ket{001}-2\ket{010}+\ket{100})+0.219677(\ket{011}-2\ket{101}+\ket{110}),\\
 \ket{\psi_3}&=&0.596008(\ket{100}-\ket{001})+0.380492(\ket{110}-\ket{011}),\\
 \ket{\psi_4}&=&-0.933013\ket{000}+0.149429(\ket{011}+\ket{101}+\ket{110})+0.25\ket{111}.
\end{eqnarray*}
It was introduced by T. V\'ertesi and N. Brunner in Ref.~\cite{PhysRevLett.108.030403}. See the reference for an analytic expression for
the amplitudes. It is constructed such that it is symmetric under permutations of the parties and invariant under partial transpose with
respect to party $3$. The last condition is sufficient for $\rho$ to be biseparable on the partition $(1,2)|3$~\cite{PhysRevA.61.062302}.
Together these conditions ensure that the state is separable along any biseparation. Therefore it is fully nondistillable. Here ``fully
nondistillable'' refers to the fact that none of the three groupings $(1,2)|3$, $(1,3)|2$ and $(2,3)|1$ of subsystems to parties is
distillable. V\'ertesi and Brunner also found that $\rho$ can be used to violate the Bell inequality 
\begin{equation}
 -13\leq\mbox{sym} [A_1 + A_1B_2 -A_2B_2 - A_1 B_1 C_1-A_2B_1C_1+A_2B_2C_2]\leq 3, \label{eq:originalbellineq}
\end{equation}
which is listed under number 5 in Ref.~\cite{Sliwa2003165}. The symbol $\mbox{sym}[X]$ denotes the symmetrization of $X$ with respect to the
three parties, e.g. $\mbox{sym}[A_1B_2]=A_1 B_2+A_1 C_2+A_2 B_1+A_2 C_1+B_1 C_2+B_2 C_1$. As $\rho$ is fully nondistillable and
nonlocal it is fully bound entangled.\\
We now use the method of homogenization described by Y. Wu and M. \zukowski in Ref.~\cite{PhysRevA.85.022119}:
By adding a constant $5$ to inequality (\ref{eq:originalbellineq}) the bounds become symmetric. Then we introduce new observables $A_0$,
$B_0$ and $C_0$ which also take the values $-1$ and $1$. Substituting the observables $A_i$ by $A_i/A_0$, $B_i$ by $B_i/B_0$ and $C_i$
by $C_i/C_0$ and factoring out $1/(A_0 B_0 C_0)$, one expands lower order correlation terms to full correlation terms. We arrive at the
inequality
\begin{eqnarray}
 &\left| \frac{1}{A_0B_0C_0} \mbox{sym} [\right.5 A_0 B_0 C_0 +A_1 B_0 C_0 + A_1B_2C_0 -A_2B_2C_0& \nonumber\\
 &\phantom{{}\left| \frac{1}{A_0B_0C_0} \mbox{sym} [\right.5 A_0 B_0 C_0{}} - A_1 B_1
C_1-A_2B_1C_1+A_2B_2C_2]\left.\vphantom{\frac{1}{A_0B_0C_0}\mbox{sym}[}\right| &\leq 8 \nonumber\\
 \Leftrightarrow\qquad&\phantom{\left| \frac{1}{A_0B_0C_0}\right.{}}| \mbox{sym} [5 A_0 B_0 C_0+A_1 B_0 C_0 + A_1B_2C_0
-A_2B_2C_0&\nonumber\\
 &\phantom{{}\left| \frac{1}{A_0B_0C_0} \mbox{sym} [\right.5 A_0 B_0 C_0{}}- A_1 B_1 C_1-A_2B_1C_1+A_2B_2C_2]| &\leq
8,\label{eq:newformbellineq}
\end{eqnarray}
which is expression H05 given in table I of Ref.~\cite{PhysRevA.85.022119}. This inequality has the required form to link to the
communication complexity problem described above. Like in Ref.~\cite{PhysRevLett.108.030403} we choose
\begin{eqnarray}
 &A_1=B_1=C_1=\left(
\begin{array}{cc}
 \cos\left(\frac{2 \pi }{9}\right) & \sin\left(\frac{2 \pi }{9}\right) \\
 \sin\left(\frac{2 \pi }{9}\right) & -\cos\left(\frac{2 \pi }{9}\right)
\end{array}
\right)\\
\mbox{and }&A_2=B_2=C_2=\left(
\begin{array}{cc}
 \sin\left(\frac{\pi }{18}\right) & -\cos\left(\frac{\pi }{18}\right) \\
 -\cos\left(\frac{\pi }{18}\right) & -\sin\left(\frac{\pi }{18}\right)
\end{array}
\right).
\end{eqnarray}
For the new observables it is sufficient to choose $A_0=B_0=C_0=\1$. With these observables we calculate the left-hand side of
(\ref{eq:newformbellineq}) using the quantum mechanical expectation values as
\begin{equation}
 S=5+3.00685=8.00685.
\end{equation}
This violation of the Bell inequality (\ref{eq:newformbellineq}) implies a quantum advantage in the quantum communication complexity task
associated with it. We write the coefficients in front of correlations $A_{x_1} B_{x_2} C_{x_3}$ in inequality (\ref{eq:newformbellineq}) as
\begin{eqnarray}
 g(x_1,x_2,x_3)&=&\phantom{\times}\left\{ 2 \left[(\delta_{x_1,x_2,x_3}+x_1+x_2+x_3) \bmod 2\right]-1\right\}\nonumber\\
&\phantom{=}&\times (1+ 4 \delta_{0,x_1,x_2,x_3}) (1-\delta_{2, (x_1 + x_2 + x_3) \bmod 3}) \prod_{i=1}^3
(1-\delta_{3,x_i}),\label{eq:gfuerh05}
\end{eqnarray}
where the symbol $\delta$ is $1$ if all subscripts are equal and $0$ otherwise. The first factor of Eq.~\ref{eq:gfuerh05} gives the sign of
the coefficient while the others define the probability distribution for $x_1$, $x_2$ and $x_3$ (see Eq.~\ref{eq:Q}). The task for the three
parties is to calculate the function
\begin{eqnarray}
 f&=&y_1 y_2 y_3 \mbox{sign}\left[g(x_1,x_2,x_3)\right]\nonumber\\
  &=&y_1 y_2 y_3 \left\{ 2 \left[(\delta_{x_1,x_2,x_3}+x_1+x_2+x_3) \bmod 2\right]-1\right\},
\end{eqnarray}
which is basically the parity of the sum of $x_1$, $x_2$, $x_3$ and $\delta_{x_1,x_2,x_3}$. As we chose $A_0=B_0=C_0=\1$ a party $i$
performs no measurement if $x_i=0$ and simply sends $y_i$. Using equations (\ref{eq:PC}) and
(\ref{eq:PQ})
we get $P_C=0.681818$ and $P_Q=0.681974$. This shows that albeit slightly, the parties still can increase the probability of success if they
share the
bound entangled state $\rho$, as compared to any classical protocol. This is striking, especially if you remind yourself that the
state $\rho$ is separable along any bipartition, i.e. it satisfies all Bell inequalities across every bipartition.
The presented task is a simple application associated with the Bell inequality (\ref{eq:originalbellineq})
the authors of Ref.~\cite{PhysRevLett.108.030403} were asking for. We note that a similar advantage can be shown using the nonlocal games
from Ref.~\cite{Silman20083796}.
\begin{acknowledgments} 
We thank Sylvia Bratzik and Dagmar Bru\ss{ }for advising us to this topic. CB acknowledges support from the European Commission,
Q-ESSENCE (No 248095) and the Austrian Science Fund (FWF): [SFB-FOCUS], [P 24621] and the doctoral programme CoQuS.
\end{acknowledgments}

\end{document}